\documentclass[12pt,dvips]{scrartcl}
\pdfoutput=1
\usepackage{amssymb}
\usepackage[pdftex]{graphicx}
\DeclareGraphicsExtensions{.pdf,.ps,.eps.gz,.ps.gz,.eps.Z}

\renewcommand{\qquad}{\hspace*{25pt}}

\newcommand{\Tr}{\mathop{\mathrm{Tr}}\nolimits}

\def\pt(#1){({\it #1\/})}

\newcommand{\eref}[1]{(\ref{#1})}
\newcommand{\sref}[1]{section~\ref{#1}}
\newcommand{\fref}[1]{figure~\ref{#1}}

\newcommand{\Fref}[1]{Figure~\ref{#1}}

\newcommand{\nonum}{\par\item[]}

\begin{document}
\title{Ensemble Inequivalence in the Ferromagnetic $p$-spin Model in Random Fields}
\author{Zsolt Bertalan\footnote{zsolt@stat.phys.titech.ac.jp}, Takehiro Kuma,\\ Yoshiki Matsuda and Hidetoshi Nishimori}

\maketitle

\begin{center}
Department of Physics, Tokyo Institute of Technology, Oh-okayama, \\Meguro-ku, Tokyo 152-8551, Japan
\end{center}
\begin{abstract}
We study the effect that randomness has on long-range interacting systems by using the ferromagnetic Ising model with $p$-body interactions in random fields. The case with $p=2$ yields a phase diagram similar to that of previously studied models and shows known features that inequivalence of the canonical and microcanonical ensembles brings with it, for example negative specific heat in a narrow region of the phase diagram. When $p>2$, however, the canonical phase diagram is completely different from the microcanonical one. The temperature does not necessarily determine the microcanonical phases uniquely, and thus the ferromagnetic and paramagnetic phases are not separated in such a region of a conventional phase diagram drawn with the temperature and field strength as the axes. Below a certain value of the external field strength, part of the ferromagnetic phase has negative specific heat. For large values of the external field strength the ergodicity is broken before the phase transition occurs for $p>2$. Moreover, for $p>2$, the Maxwell construction cannot be derived in a consistent manner and therefore, in contrast to previous cases with negative specific heat, the Maxwell construction does not bridge the gap between the ensembles. 
\end{abstract}

\section{Introduction}

In recent years much effort has been devoted to the study of the inequivalence of ensembles of systems with long-range interactions \cite{hertel,barre,barre2,mukamel,campa,mukamel2}. Systems where the interaction decays as $\sim 1/r^{\alpha}$, with $\alpha$ less than or equal to the dimension of the system, are said to be of long-range. Such systems are non-additive, meaning that interface energies between macroscopic systems are not negligible compared to the bulk energies. Non-additive, long-range interacting systems are found in many areas of physics, prominent examples being self-gravitating systems \cite{antonov, lyb-wood} and spin systems of mean field nature \cite{barre, mukamel,ispolatov}. Some characteristics of long-range interactions have also been observed in driven systems with only local stochastic dynamics \cite{mukamel_ABC}. An important feature of systems with long-range interactions is that thermodynamical properties can be dependent on from which ensemble they were derived \cite{campa}. For example, the specific heat in the canonical ensemble is always positive, but may become negative in the microcanonical ensemble \cite{lyb-wood,lyb-thir}. For more features, intricacies and examples of long-range systems we refer the reader to \cite{campa,dauxois}.

It has been conjectured in \cite{barre} that a necessary condition for ensemble inequivalence to occur in statistical long-range interacting systems is that the system undergoes a first-order phase transition. This statement has not yet been proven generally, but has been put on a solid basis for many cases in \cite{ellis,bouchet}.

A paradigmatic system with long-range interactions is the infinite-range $p$-spin model, also known as the Ising model with infinite-range $p$-body interactions. Notable variants are the Blume-Capel \cite{BC} and the Blume-Emery-Griffiths \cite{BEG} models in the infinite-range limit , in which $p=2$ and a spin can take the values $S_i=1,0,-1$. It was shown in \cite{barre} and \cite{mukamel} that the Blume-Capel model and a model with infinite-range and short-range interactions exhibit ensemble inequivalence in the region of their phase diagram where first-order transitions occur. Also, a generalized XY model with two-body ($p=2$) and four-body $(p=4)$ was studied in the microcanonical ensemble in \cite{buyl-bouchet}.

It has not been studied, however, what would happen if a long-range interacting system has randomness. This problem is of interest, because it is the target of extensive studies in the field of spin glasses \cite{nishimori,derrida,gross}. Almost all calculations so far have been done in the canonical ensemble for spin glasses with many-body interactions. It should be necessary to check if any different results may be derived in the microcanonical ensemble.

It is quite difficult to analyze spin glass systems with many-body interactions using the microcanonical ensemble. Even the analysis in the canonical ensemble is plagued by the complications of replica symmetry breaking \cite{nishimori}, and we will have to develop ingenious methods to treat systems with randomness in the context of the microcanonical ensemble. We therefore set our goal more modest in the present paper and study the random-field Ising model with $p$-body interactions in the hope that some general features of the interplay between many-body interactions, randomness and first-order transitions may manifest themselves, leading to non-trivial consequences in ensemble inequivalence already in this relatively simple random system.

This paper is organized as follows. In \sref{sec:model} we introduce the model and derive the thermodynamic potentials, the free energy in the canonical ensemble and the entropy in the microcanonical ensemble, and discuss some of their general features. In \sref{sec:comparison} we compare the phase diagrams of both ensembles and discuss their differences. Our study is concluded in \sref{sec:conclusio}. 

\section{The Model and General Considerations}\label{sec:model}

In this section we introduce the ferromagnetic $p$-spin model (= Ising model with $p$-body interactions) in random fields, derive the thermodynamical potentials in the canonical and microcanonical ensembles for later comparison in \sref{sec:comparison}. 

The ferromagnetic $p$-spin model in external fields describes a system of $N$ spins, in our case Ising spins ($S_i=\pm 1$), with infinite-range $p$-body interactions and each site is subject to a random external field. It is defined by the Hamiltonian,
\begin{eqnarray}\label{eqn:HAM}
 H=-N\left(\frac{J}{N}\sum_i S_i\right)^p-\sum_i h_iS_i,
\end{eqnarray}
with $J>0$. Throughout this study we will set $J=1$ without loss of generality. The external field is assumed to obey a quenched bimodal distribution given by
\begin{eqnarray}\label{eqn:h_distr}
 P(h_i)=\frac{1}{2}\delta(h_i-h_0)+\frac{1}{2}\delta(h_i+h_0).
\end{eqnarray}

\subsection{Canonical Solution - Free Energy}

The partition function can be calculated in the standard manner by the introduction two auxiliary fields \cite{nishimori}
\begin{eqnarray}
 Z &=& \Tr \int dm\ \delta\left(Nm -\sum_iS_i\right)\exp\left(\beta Nm^p+\beta\sum_ih_iS_i\right) \nonumber\\
 &=&\Tr \int d\bar m\ dm\ \exp\left(\beta Nm^p-Nm\bar m +\bar m\sum_iS_i +\beta\sum_ih_iS_i\right) \nonumber\\
 &=&\int d\bar m\ dm\ \exp\left({N\beta m^p - N\bar{m}m+\sum_i \ln(2\cosh(\bar m +\beta h_i))}\right),
\end{eqnarray}
where $\beta$ is the inverse temperature. Evaluating this integral by the saddle point method, together with the saddle-point equations
\begin{eqnarray}
 \bar m &=& \beta p m^{p-1}\\
 m&=&\frac{1}{N}\sum_i^N \tanh(\bar m +\beta h_i)\label{eqn:sp_2},
\end{eqnarray} 
leads to the mean-field free energy per spin,
\begin{eqnarray}
 \beta f(m,\beta) &=& \beta (p-1) m^p -\frac{1}{N} \sum_i \ln 2\cosh\beta(pm^{p-1} + h_i) \nonumber \\
 &=&\beta (p-1) m^p -\frac{1}{2} \ln 2\cosh\beta(pm^{p-1}+ h_0)\nonumber\\
 && -\frac{1}{2} \ln 2 \cosh\beta(pm^{p-1} - h_0)\label{eqn:cn_f},
\end{eqnarray}
where $m$ obeys the self-consistent equation 
\begin{eqnarray}\label{eqn:sc_cn}
 m=\frac{1}{2} \tanh\beta(pm^{p-1} + h_0)+\frac{1}{2} \tanh\beta(pm^{p-1} - h_0).
\end{eqnarray}
We have used the self-averaging property \cite{nishimori} (meaning that $\sum_ig(h_i)/N$ is equal to the configurational average of $g(h_i)$ for sufficiently large $N$, where $g$ is some physical quantity) to rewrite the site average in the first line of \eref{eqn:cn_f} by the randomness average in the second line.
The free energy \eref{eqn:cn_f} shows standard behavior of first- and second-order transitions depending on the values of $p$ and $h_0$. Details of the canonical phase diagram will be elucidated in the next section.


\subsection{Microcanonical Solution - Entropy}

\begin{figure}[ht]
 \centering
 \includegraphics[width=\textwidth]{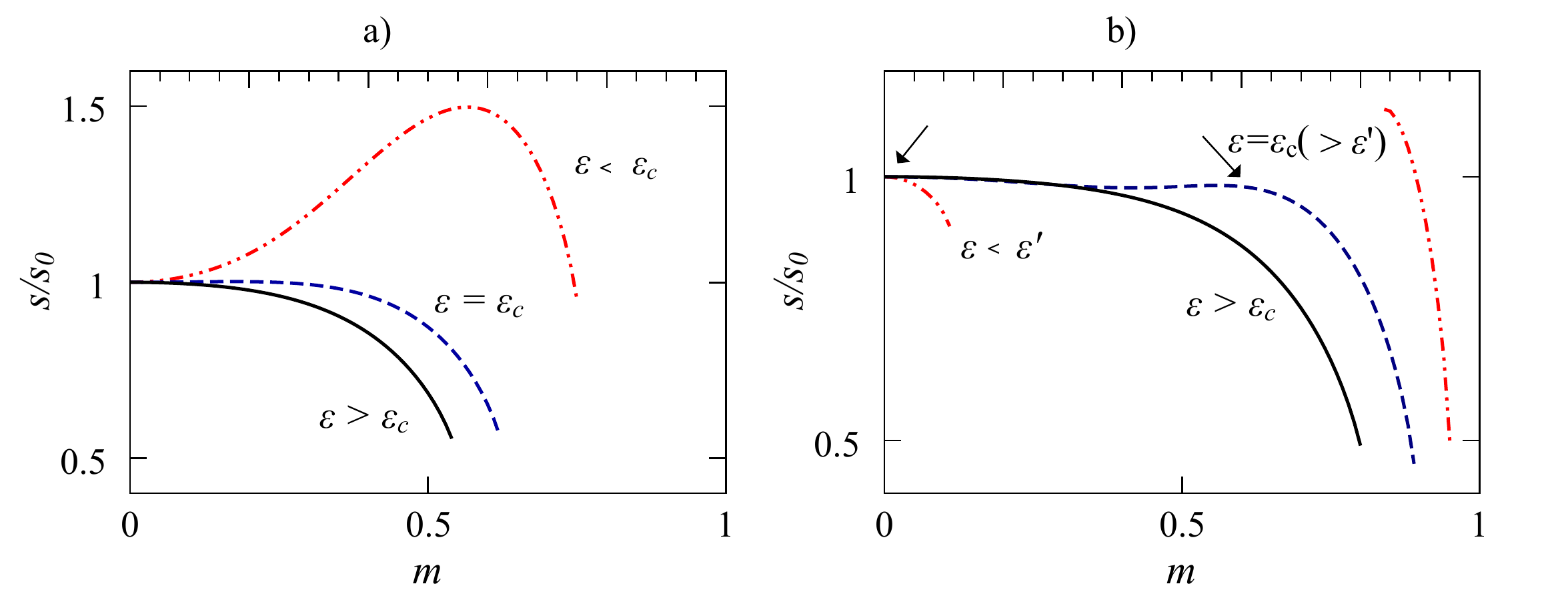}
 \caption{Microcanonical entropy per spin versus magnetization per spin, scaled to $s_0$, its value at $m=0$, for $p=2$, at external field strengths a) $h_0=0.55$ and b) $h_0=0.95$. In b) it is seen that the entropy for this case with a first-order transition is not a continuous function of the magnetization for energies lower than $\epsilon'$, which lies below the critical energy for the phase transition, $\epsilon'<\epsilon_c$. The two maxima at $\epsilon=\epsilon_c$, indicated by arrows, have equal height. }\label{fig:mc_entropies}
\end{figure}

\begin{figure}[ht]
 \centering
 \includegraphics[width=\textwidth]{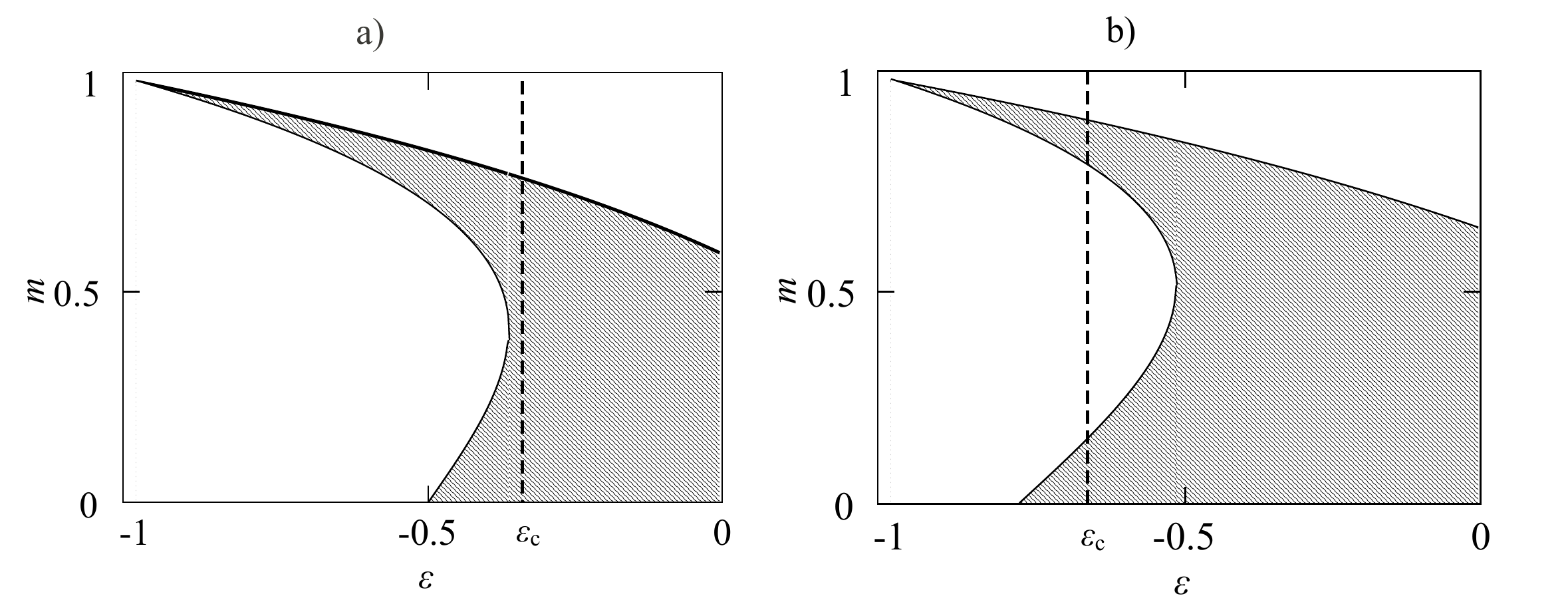} 
 \caption{The allowed magnetization per spin (shaded) versus the energy per spin for $p=3$ at a)  $h_0=0.5$ and b) $h_0=0.782$. At low energy the allowed region of magnetization splits into two parts. The system in one region of magnetization cannot evolve to the other region under the microcanonical constraint of constant energy. The vertical dashed line shows at which energy $\epsilon_c$ the microcanonical phase transition occurs. For $h_0=0.5$ the phase transition occurs before the ergodicity breaks, while for $h_0=0.782$ the ergodicity breaks before the phase transition.}\label{fig:p3_me}
\end{figure}

The microcanonical entropy per spin is calculated from $s=\ln \Omega/N$, where $\Omega$ is the sum of states for given values of $m$ (magnetization per spin) and $\epsilon$ (energy per spin). The calculations are straightforward, as sketched in \ref{sec:app_entropy}, and the result is
\begin{eqnarray}\label{eqn:mc_s}
 s(m,\epsilon ,h_0) &=&-\left(\frac{1+m}{4}-\frac{\epsilon+m^p}{4 h_0}\right)\ln\left(\frac{1+m}{4}-\frac{\epsilon+m^p}{4 h_0}\right) \nonumber \\ 
 &&- \left(\frac{1+m}{4}+\frac{\epsilon+m^p}{4h_0}\right)\ln\left(\frac{1+m}{4}+\frac{\epsilon+m^p}{4h_0}\right) \nonumber \\
 && -\left(\frac{1-m}{4}-\frac{\epsilon+m^p}{4h_0}\right)\ln\left(\frac{1-m}{4}-\frac{\epsilon+m^p}{4h_0}\right) \nonumber \\ 
 && - \left(\frac{1-m}{4}+\frac{\epsilon+m^p}{4h_0}\right)\ln\left(\frac{1-m}{4}+\frac{\epsilon+m^p}{4h_0}\right)-\ln 2.
\end{eqnarray}
 In the microcanonical ensemble, the basic strategy, corresponding to the free-energy minimization in the canonical ensemble, is the maximization of the entropy. See \fref{fig:mc_entropies} for the entropy as a function of the magnetization at different energies for $p=2$. We have a paramagnetic phase, when the entropy at $m=0$ is the global maximum. When the transition is second order as in \fref{fig:mc_entropies} a), the entropy flattens around $m=0$ as the energy decreases and develops a maximum at a non-zero value of the magnetization, while the region around $m=0$ becomes a minimum, which implies an instability of the paramagnetic phase. In the case of a first-order transition (\fref{fig:mc_entropies} b)), as the energy decreases, the entropy develops a second maximum, and at a critical value $\epsilon_c$ the two maxima are of equal height. As the energy decreases further, this second maximum dominates. At some energy $\epsilon'$ the entropy becomes a discontinuous function of the magnetization and the ergodicity is broken: Intermediate values of the magnetization are not allowed, and consequently, the system in the low-$m$ region cannot evolve dynamically to the high-$m$ region. As a consequence, the metastable state with $m=0$ has a divergent relaxation time \cite{mukamel}. The energy, where ergodicity is broken, lies below the critical energy $\epsilon'<\epsilon_c$ for $p=2$, but for $p\geq 3$ we have $\epsilon'>\epsilon_c$ for high $h_0$, while for low $h_0$ the ergodicity breaks after the phase transition takes place. \Fref{fig:p3_me} shows this property as the restriction on the accessible region in the $(m,\epsilon)$-plane, the non-convexity of the region being closely related to the ergodicity-breaking. This unusual behavior is not unique to this model, but has been observed in other systems with ensemble inequivalence \cite{barre,mukamel,campa}. 

\section{Canonical and Microcanonical Phase Diagrams} \label{sec:comparison}

In this section we investigate the thermodynamic properties of the model (1) introduced in the previous section. We compare and discuss the phase diagrams for the cases $p=2$ and $p>2$ in the canonical and microcanonical ensembles and show explicitly that in certain intervals of $h_0$ and $\epsilon$ the specific heat of the ferromagnetic phase is negative in the microcanonical ensemble, which is a clear indication of ensemble inequivalence. 

\subsection{$p=2$}\label{sec:p2}

\begin{figure}[ht]
 \centering
 \includegraphics[width=\textwidth]{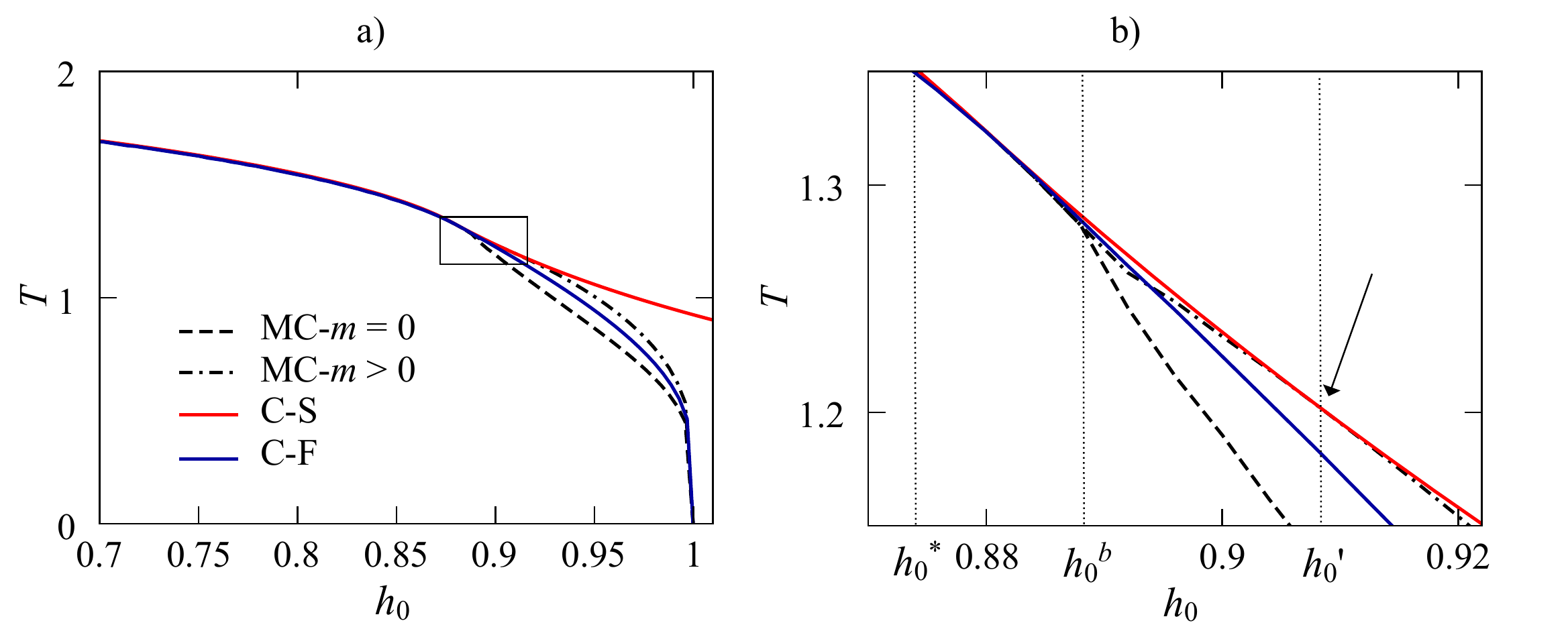} 
 \caption{a) The canonical and microcanonical ($h_0,T$) phase diagrams at $p=2$. b) Details of the phase diagram, showing the region around the tricritical points, bounded by the box in a). The arrow points where the canonical spinodal line (C-S, red) touches the microcanonical-$m>0$ line (MC-$m>0$, dash-dotted). For values of the external field strength $h_0^*<h_0<h_0'$, the specific heat is negative for energies lower than the transition energy up to a certain energy value. At $h_0^b $ the temperature develops a jump at the transition energy. The red line touches the dash-dotted line as indicated by an arrow.}\label{fig:p2_pd}
\end{figure}

The phase diagram of the model with $p=2$, as shown in \fref{fig:p2_pd} a), is qualitatively the same as the phase diagram of the models discussed in \cite{barre,mukamel}. For small $h_0$ there is a continuous (second-order) transition in the canonical ensemble up to the canonical tricritical point, beyond which the transition becomes of first order \cite{aharony}. In the microcanonical ensemble the continuous transition coincides, for small $h_0$, with the canonical one, but persists to a higher value of $h_0$ up to the microcanonical tricritical point. The first-order transition in the microcanonical ensemble manifests itself as a jump in the temperature and is represented by two lines in the $(h_0,T)$ plane, the MC-$m>0$ and MC-$m=0$ lines in \fref{fig:p2_pd} a). The reason for this jump is that the entropy has two maxima, one at zero magnetization and one at finite magnetization indicated by the arrows in \fref{fig:mc_entropies} b). Since the temperature is calculated from the relation $1/T=(\partial s/\partial \epsilon) (\epsilon,m^*)$ (see \eref{eqn:beta} in \ref{sec:app_entropy}), and there are two values for $m^*$, zero and another finite value, there is a temperature jump in the phase diagram due to $(\partial s/\partial \epsilon) (\epsilon,m^*=0)\neq (\partial s/\partial \epsilon) (\epsilon,m^*\neq0)$ for $h_0>h_0^b$. Note that the upper microcanonical first-order line in the phase diagram in \fref{fig:p2_pd} corresponds to $m^*\neq 0$ (MC-$m>0$) and the lower curve to $m^*=0$ (MC-$m=0$). Between those two lines, the phase of the system is not determined uniquely by the temperature and thus the two phases, ferromagnetic and paramagnetic, are not completely separated in the $(h_0,T)$ plane. 

\begin{figure}[htb]
 \centering
 \includegraphics[width=\textwidth]{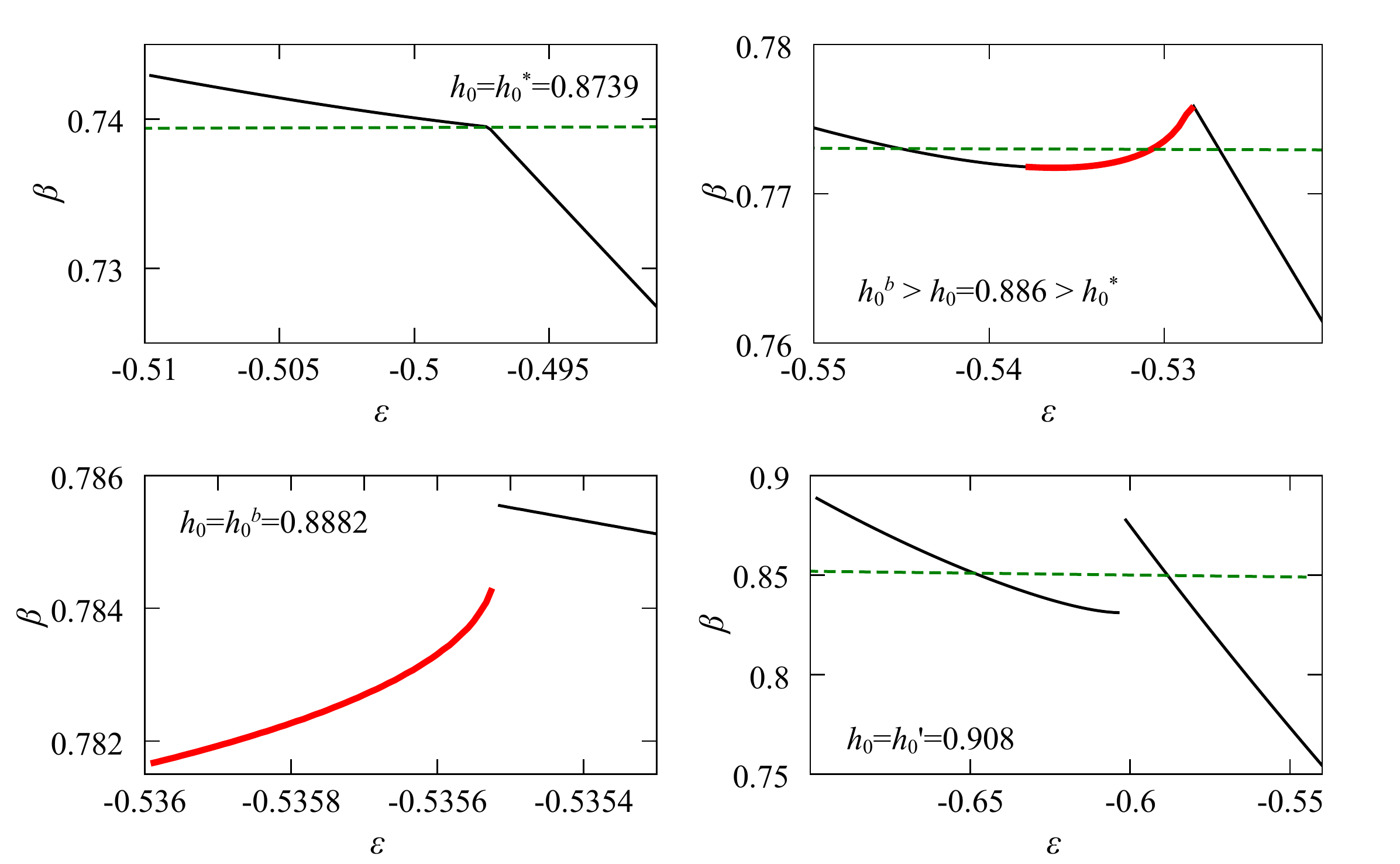} 
 \caption{Inverse temperature versus energy per spin for $p=2$. The thick, red line indicates negative specific heat, because the temperature increases as the energy decreases. The thick green dashed lines indicate the canonical transition temperatures, coinciding with the Maxwell construction.}\label{fig:p2_cv}
\end{figure}

\Fref{fig:p2_cv} shows the inverse temperature versus the energy per spin for several important values of $h_0$ in the region of the first-order transition. For $h_0> h_0^*= 0.8739$, there is a small region where the specific heat is negative, shown as thick, red lines, around the transition energy. As $h_0$ grows, the temperature develops a jump at $h_0^b\approx 0.8882$, and at $h_0'=0.908$ the negative specific heat region disappears. Thus, the region in the $(h_0,T)$ plane where the system has negative specific heat is very narrow, $h_0^*<h_0<h_0'$, as depicted in \fref{fig:p2_pd} b). 

A close look at \fref{fig:p2_pd} b) reveals that the canonical spinodal line (shown in red) touches the upper microcanonical line (shown as dash-dotted) \cite{bouchet} at $h_0=h_0'(=0.908)$, indicated by an arrow, where the specific heat becomes positive again. The dashed horizontal lines in \fref{fig:p2_cv} show the temperature where the canonical phase transition takes place. They can be obtained by the Maxwell construction \cite{reichl} which was generalized to microcanonical systems in \cite{chavanis}.  


\subsection{$p=3$} \label{ssec:p3}

The microcanonical phase diagram changes drastically when many-body interactions come into play.

\begin{figure}[htb]
 \centering
 \includegraphics[width=\textwidth]{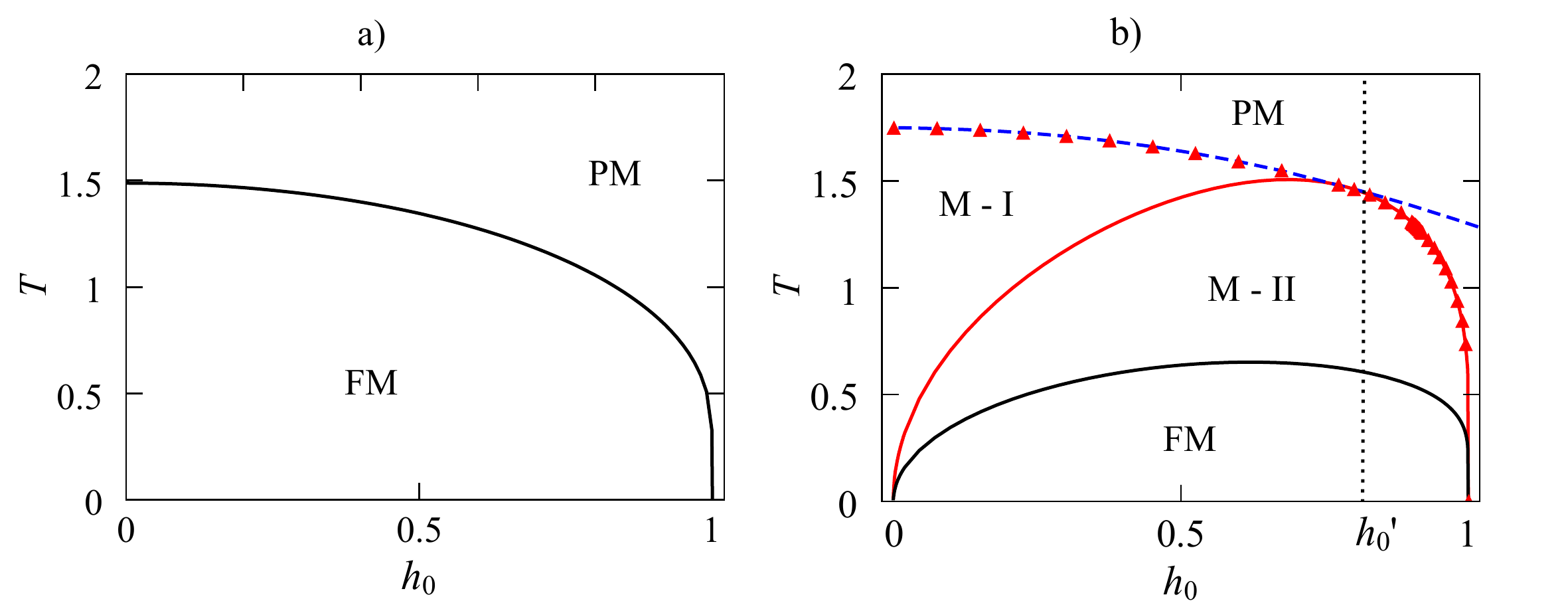} 
 \caption{The canonical phase diagram a) shows standard behavior: the ferromagnetic (FM) and paramagnetic (PM) phases are separated by a first-order transition line. The microcanonical phase diagram b) has a richer structure. Between the paramagnetic (PM) and the ferromagnetic (FM) phases lie regions where both phases coexist. In M - I the ferromagnetic solutions with high energy have negative specific heat, while in M - II all ferromagnetic solutions have positive specific heat. The phase M - I disappears for $h_0>h_0'$. The dashed blue line is the canonical spinodal line.  }\label{fig:p3_pd}
\end{figure}

 \Fref{fig:p3_pd} a) shows the canonical phase diagram for $p=3$. Here, nothing unexpected happens, the phases, ferromagnetic and paramagnetic, are cleanly separated by a first order transition. The microcanonical phase diagram as shown in \fref{fig:p3_pd} b) exhibits a far richer structure. The paramagnetic solution extends almost over the whole $(h_0,T)$ plane, except in a narrow region upper-bounded by the black line. Below that line, a ferromagnetic phase (FM) exists. However, the ferromagnetic solution extends up to the temperature indicated by red triangles. The blue dashed line is the $canonical$ spinodal line. In regions M - I and M - II between the red triangles and the black line, the ferromagnetic and paramagnetic solutions coexist, with the ferromagnetic solution having positive specific heat in region M - II, between the red and black lines. In the region M - I between the red line and red triangles, the ferromagnetic solution with high energy has negative specific heat while the low energy solution retains positive specific heat. At $h_0=h_0'$ the region of negative specific heat disappears and for $h_0>h_0'$ all stable ferromagnetic solutions have positive specific heat. 
     
\begin{figure}[htb]
 \centering
 \includegraphics[width=\textwidth]{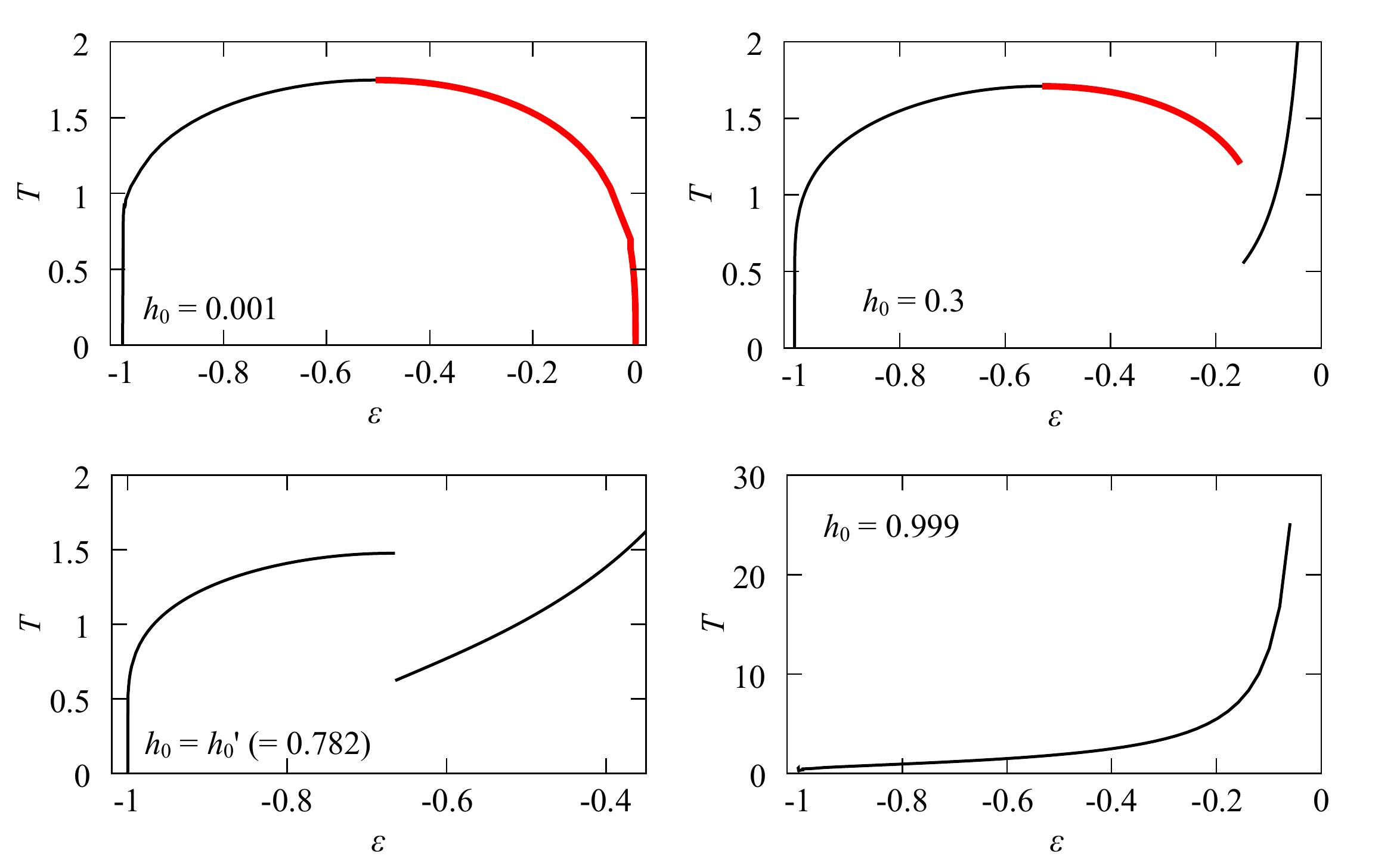} 
 \caption{At very low field strengths the energy is almost entirely determined by the magnetization, and the ferromagnetic solution dominates. At $h_0=0.001$ the paramagnetic solution is not shown. At high energies the specific heat is negative as indicated by a thick red line. As the field strength increases, the randomness destabilizes the ferromagnetic solution at  high energies. The high energy branch for $h_0=0.3$ and $h_0=0.782$ represents the paramagnetic solution. At $h_0=h_0'(=0.782)$ the whole of the negative-specific-heat solution is metastable. The ferromagnetic solution destabilizes completely at $h_0=1$.}\label{fig:p3_cv}
\end{figure}

 To understand this behavior in more detail, we have to look at caloric curves $T(\epsilon)$ at various values of the field strength. In \fref{fig:p3_cv} we show the temperature versus the energy for field strengths $h_0=0.001$, $h_0=0.3$, $h_0=0.782$ and $h_0=0.999$. At very low $h_0 (=0.001)$ the energy \eref{eqn:eperspin} is almost exclusively dominated by the magnetization and the paramagnetic solution exists only in a very narrow energy region (not shown). A noteworthy feature of the ferromagnetic solution is that at high energies it has negative specific heat (shown as thick red lines), where the temperature increases as the energy is lowered.  As the field strength increases, the effect of randomness becomes more important and the ferromagnetic solution becomes destabilized in the high energy region. At $h_0=h_0(=0.782)$ the region where the ferromagnetic solution has negative specific heat becomes metastable. At high field strengths, very close to one, almost the whole ferromagnetic solution is metastable in the sense described in the following section, remaining only at the lowest energy values. At $h_0>1$, the randomness prevents any ferromagnetic order. 

\subsection{Stability Analysis for $p=3$}
\begin{figure}
\centering
\includegraphics[width=\textwidth]{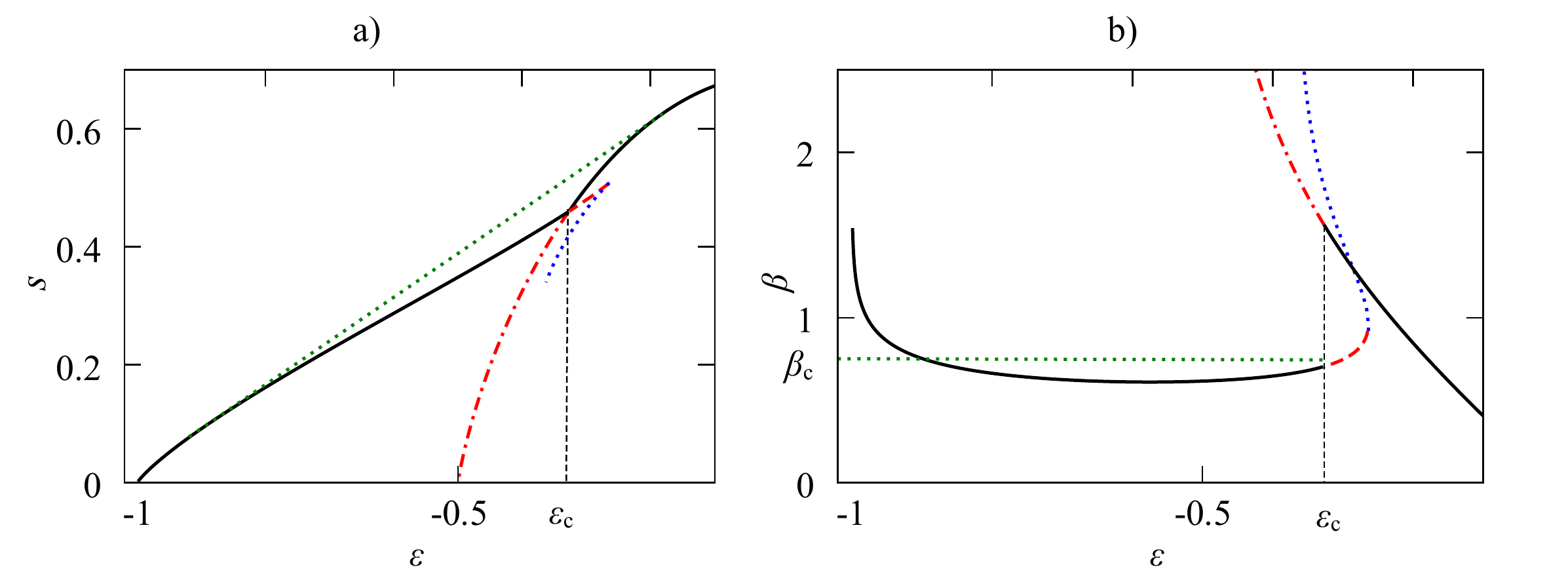} 
\caption{ a) The different branches of the entropy for $h_0=0.5$. The black line is the equilibrium entropy, while the red dash-dotted line is the paramagnetic metastable solution and the red dashed line the ferromagnetic metastable solution. The unstable solution is the blue dotted line. The green dotted, oblique line is the concave-cover construction. b) The corresponding caloric curves. Below $\epsilon_c$ the ferromagnetic solution is stable (black) and the paramagnetic solution is metastable (red dash-dotted), while above $\epsilon_c$ the paramagnetic solution is stable (black) and the ferromagnetic solution is metastable (red dashed). The unstable branch is shown blue dotted.}\label{fig:p3_se}
\end{figure}

\begin{figure}[htb]
 \centering
 \includegraphics[width=\textwidth]{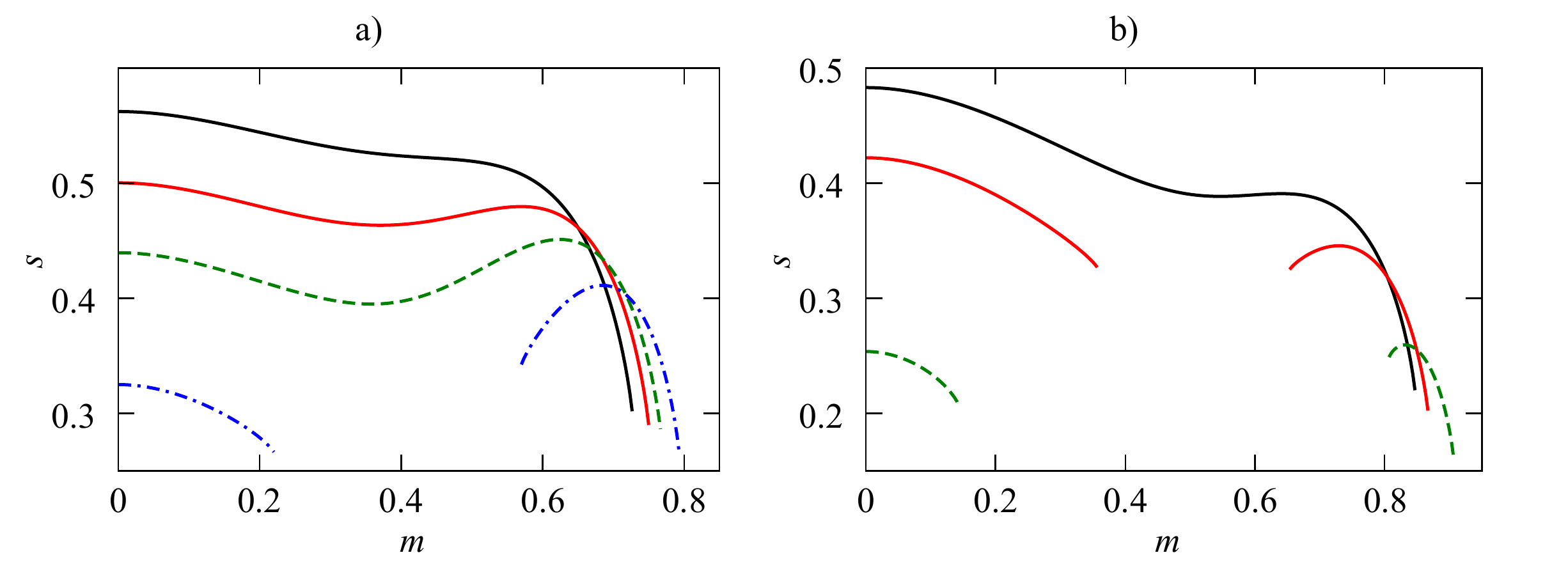} 
 \caption{A typical example of the entropy per spin versus the magnetization per spin for $p=3$. a) Low field strength $h_0=0.5$, at energies $\epsilon=-0.25$ (black), $\epsilon=-0.3$ (red), $\epsilon=-0.34$ (green dashed)  and $\epsilon=-0.4$ (blue dash-dotted) b) High field strength $h_0=0.782$ at energies $\epsilon=-0.49$ (black), $\epsilon=-0.55$ (red) and $\epsilon=-0.6$ (green dashed). For $h_0=0.782$, the ergodicity is broken before the maximum at $m>0$ reaches the same height as the maximum at $m=0$. The unstable solution corresponds to the minimum of the entropy. }\label{fig:p3_ms}
\end{figure}

To clarify what we mean by stable, unstable and metastable solutions, let us look at the entropy per spin as a function of the energy per spin. As stated earlier, the entropy has a maximum at zero magnetization. As the energy decreases, a second maximum emerges. These two branches of the entropy are in direct competition as shown in \fref{fig:p3_se} a). Above the critical energy $\epsilon_c$ the maximum at $m=0$ is higher, and thus the paramagnetic branch wins, shown in black, while the ferromagnetic branch lies at lower entropy, shown red dashed. Below $\epsilon_c$, the ferromagnetic branch wins (black), while the paramagnetic branch lies at lower entropy, shown red dash-dotted. Note that there is a sharp cusp at the transition, and the entropy is not smooth. This is a direct consequence of the fact that the unstable solutions (blue line) do not mediate the phase transition in bridging the gap between the metastable ferromagnetic and paramagnetic solutions. Thus, they can not smoothen the entropy around the transition point as is often observed in systems with long-range interactions including the van der Waals gas/liquid. The unstable solution corresponds to the minimum of the entropy with respect to the magnetization (see \fref{fig:p3_ms}). The caloric curves corresponding to the entropy branches in \ref{fig:p3_se} a) are shown in \fref{fig:p3_se} b). The paramagnetic solution (black) is stable above $\epsilon_c$ and metastable below (red dash dotted), while the situation for the ferromagnetic solution is the exact opposite: stable below (black) and metastable above (red dashed) $\epsilon_c$. As the ergodicity is broken, the unstable solution (blue dashed) disappears. However, at high field strengths the ergodicity breaks before the phase transition takes place (\fref{fig:p3_ms} b)). Other models with many-body interactions, but without randomness, also show ergodicity breaking before the phase transition occurs \cite{buyl-bouchet}.

The Maxwell construction cannot be used to bridge the gap between the microcanonical and canonical ensembles. The conventional Maxwell construction yields the canonical transition temperature from the caloric curve $\beta(\epsilon)$, while the microcanonical Maxwell construction \cite{campa,chavanis} yields the transition energy and the corresponding canonical transition temperature. However, both procedures rely on the fact that the unstable solution interpolates the gap between the paramagnetic metastable and ferromagnetic metastable solutions. In our case, the Maxwell construction is impossible. For low $h_0$, the unstable solutions do not bridge the metastable solutions. As shown in \fref{fig:p3_se} b), the unstable solution (blue dotted) does not touch the paramagnetic metastable solution (red dash-dotted). On the other hand, for high $h_0$, the unstable solution disappears even before the phase transition occurs due to ergodicity breaking. For more details on the Maxwell construction we refer the reader to the literature \cite{campa,reichl,chavanis}. 

It is useful to note here that the microcanonical entropy and the corresponding caloric curve in \fref{fig:p3_se} can be reproduced from the generalized free energy \eref{eqn:cn_f}. Details are delegated to \ref{sec:mcentfromgf}.


Qualitatively similar results have been found for $p>3$.

\subsection{The Limit $p\rightarrow\infty$}\label{ssec:pinf}
\begin{figure}
\centering
\includegraphics[width=\textwidth]{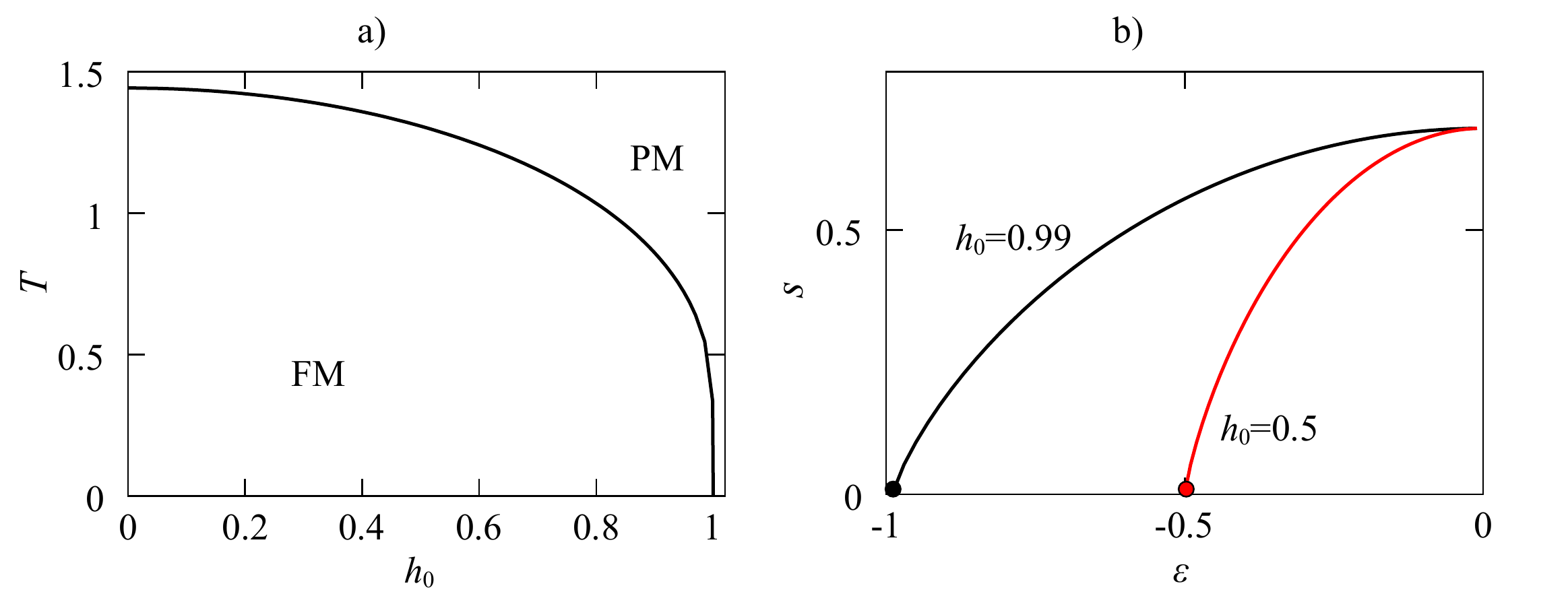} 
\caption{a) Canonical phase diagram for $p=\infty$. The ferromagnetic (FM) and paramagnetic (PM) phases are clearly separated. b) Microcanonical entropy for $p=\infty$ at external fields $h_0=0.99$ (black) and $h_0=0.5$ (red). The system is paramagnetic at all temperatures, because $\epsilon=0$ ($m=0$) has always the largest entropy.}\label{fig:pp_pd}
\end{figure}

It is important to see what happens when we let $p\rightarrow\infty$, because this limit plays important roles in some problems related to spin glasses \cite{nishimori,derrida,gross}. The canonical phase diagram can be obtained analytically. If we take the limit $p\rightarrow\infty$ in equations \eref{eqn:cn_f} and \eref{eqn:sc_cn} and assuming $|m|<1$, we see that $m=0$ is the only solution for $|m|<1$ and the free energy for the paramagnetic phase is simply $f_P=-\ln(2\cosh(\beta h_0))/\beta$. In the same limit, $p\rightarrow\infty$, $m=1$ is also a solution of \eref{eqn:sc_cn} and the ferromagnetic free energy is $f_F=-1$ irrespective of the temperature. Another solution with $|m|=1$, i.e. $m=-1$, is not allowed for $p$ odd as one sees easily from \eref{eqn:sc_cn}. For $p$ even, $m=-1$ is also a solution but with the same free energy as $m=1$. At the transition point the free energy should be continuous $f_P=f_F$, which leads to the condition for the critical line
\begin{eqnarray}
 2\cosh\beta h_0=e^{\beta}.
\end{eqnarray}
The canonical phase diagram is shown in \fref{fig:pp_pd} a).

In the microcanonical case, letting $p\rightarrow\infty$ in \eref{eqn:mc_s} and assuming $|m|<1$ leads to the conclusion that the only stable solution is at $m=0$. With \eref{eqn:beta} the paramagnetic solution leads to
 \begin{eqnarray}
 \epsilon=-h_0\tanh\beta h_0.
\end{eqnarray}
The ferromagnetic solution, $|m|=1$, yields $\epsilon=-1$ and the temperature cannot be defined since the entropy has the unique value $s=0$ at the sole allowed energy $\epsilon=-1$, which forbids us to evaluate $\partial s/\partial\epsilon$. However, at zero temperature, the entropy of the paramagnetic solution has the same value, $s=0$, as the ferromagnetic entropy. 

\begin{figure}
\centering
\includegraphics[width=\textwidth]{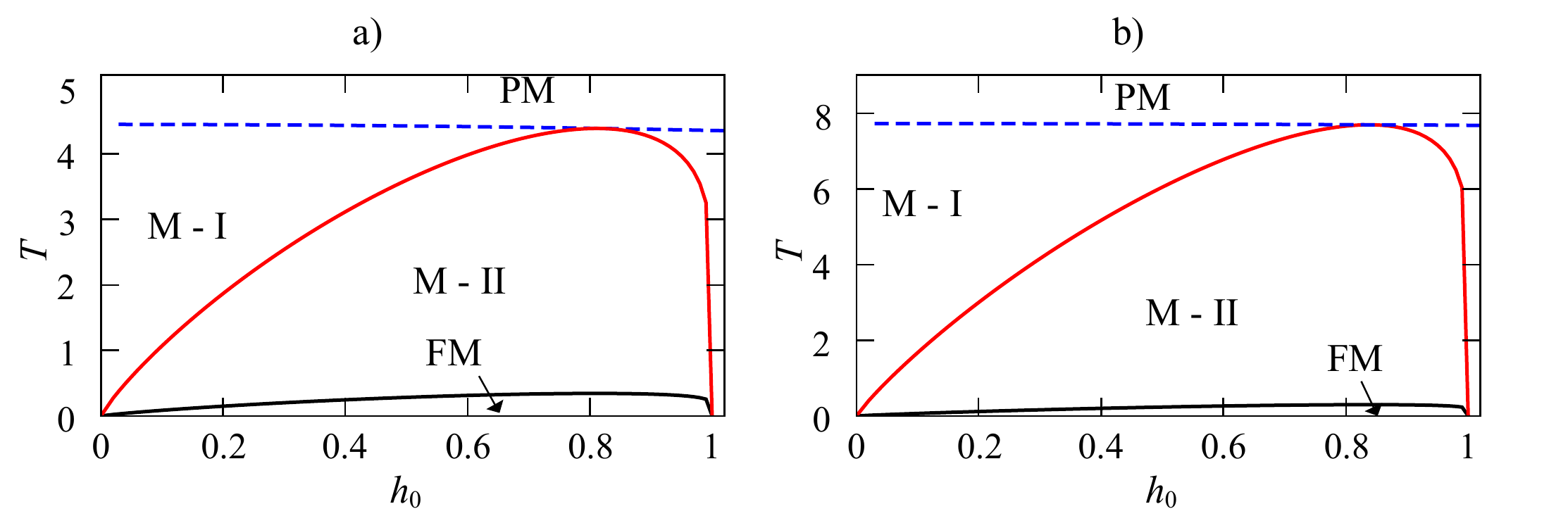} 
\caption{Microcanonical phase diagram for a) $p=13$ and b) $p=26$. The regions where the pure ferromagnetic and paramagnetic phases exist shrink with growing $p$, while the mixed phases, M - I and M - II, blow up. }\label{fig:p_inf_pd}
\end{figure}

 To better understand this behavior, we investigate the dependence of the microcanonical phases for large $p$ in \fref{fig:p_inf_pd} a) $p=13$ and b) $p=26$. We see that the ferromagnetic region becomes smaller as $p$ increases, while the temperature region where the ferromagnetic phase coexists with the paramagnetic phase extends to higher and higher temperatures and the whole region in the phase diagram tends to be M - II. Consequently, the limit $p \rightarrow\infty$ gives different results than when the calculations are done with $p=\infty$.
 

\section{Summary and Discussion}\label{sec:conclusio}
We have performed a comparison of the microcanonical and canonical phase diagrams of the ferromagnetic $p$-spin model with bimodally distributed random external fields and with infinite-range interactions.

For $p=2$, both ensembles coincide in the region of low values of the external-field strength $h_0$, where the transition is of second order up to the canonical tricritical point. From there on the phase diagrams disagree. The microcanonical second-order transition persists, off the canonical phase boundary, to a higher value of $h_0$ up to the microcanonical tricritical point. The first-order transition from the paramagnetic to the ferromagnetic state is accompanied by a temperature jump. The specific heat of the ferromagnetic phase, in the microcanonical ensemble, is negative in a very small interval of the external-field strength and only around the transition energy. All in all, the microcanonical phase diagram of the infinite-range model with random fields for $p=2$ is comparable to that of previously studied models \cite{barre,mukamel}. 

For the case $p>2$, only a first-order transition exists in both ensembles. The two canonical phases are cleanly separated from each other, with the ferromagnetic phase lying, as expected, at lower temperatures than the paramagnetic phase. The microcanonical case, however, gives completely different behavior. The paramagnetic phase extends down to low temperatures, while the ferromagnetic phase extends up to the canonical spinodal line below a certain value of the field strength $h_0'$ and up to some other temperature, below the canonical spinodal line. The mixed phase below $h_0'$ splits again into two regions, one where the ferromagnetic solution has positive specific heat and another region where the specific heat may take negative values. 

Also, no conventional Maxwell construction is possible for $p>2$, in contrast to the case of $p=2$, where it could reconcile the microcanonical and canonical phase diagrams. The reason for this is that, while for $p=2$ the metastable solutions are joined smoothly by the unstable solutions, for $p=3$ the unstable solution does not touch the paramagnetic metastable solution. For high values of the external field, the ergodicity is broken before the phase transition occurs, and the unstable solution does not even exist there. 
%

In the limit $p\rightarrow\infty$ the canonical phase diagram can be obtained analytically. It is qualitatively comparable to the case of $p=3$. In the microcanonical ensemble, however, only the paramagnetic phase exists at finite temperatures and the temperature of the ferromagnetic phase is ill-defined for $p=\infty$, but for $p$ large (i.e. in the limit $p\rightarrow\infty$), all temperatures have both ferromagnetic and paramagnetic solutions. 

We have therefore found that subjecting the many-body Ising model to random external fields yields many interesting and unexpected phenomena. Some already known features, like ergodicity breaking and negative specific heat, have shown up also in this model. New features of the system emerge when $p>2$, such the impossibility of the conventional Maxwell construction or the breaking of ergodicity before or after the phase transition occurs, depending on the external control parameter. Clearly, these facts demand more detailed studies of the effect of randomness in long-range interacting systems. New microcanonical methods have to be developed to tackle the emerging difficulties that disorder brings to this system, especially in the context of spin glasses, and studies in this direction are underway.

\section*{Acknowledgements}
We thank the CREST project, JST for financial support. We thank David Mukamel, Stefano Ruffo and Thierry Dauxois for encouragement.

\appendix

\section{Microcanonical Entropy}\label{sec:app_entropy}
Here, we derive explicitly the entropy, \eref{eqn:mc_s}, from the sum of states and analyze its derivatives.

Let us denote the number of sites with the field quenched to $h_i(=h_0$ or $-h_0)$ and spin $S_i$ as $N_{S_ih_i}$. Then there are four possible denotations, since the field can be $+h_0$ or $-h_0$ and the spin can be $\pm 1$:
 $N_{++}$ for spins $+1$ at sites with $h_i=+h_0$, $N_{+-}$ for spins $-1$ at sites with $h_i=+h_0$, and so on. Since half of the $h_i$ are $+h_0$ and the other half $-h_0$, the number of states is given by
 \begin{eqnarray}
\Omega = \frac{\left({N}/{2}\right)!}{N_{++}!N_{-+}}\frac{\left({N}/{2}\right)!}{!N_{+-}!N_{--}!}.
\end{eqnarray}
The energy per spin is calculated from \eref{eqn:HAM} as 
\begin{eqnarray}
\epsilon &=& = -\left(\frac{1}{N}\sum_i S_i\right)^p -h_0\frac{N_{++}-N_{+-}}{N}+h_0\frac{N_{-+}-N_{--}}{N}\nonumber \\
&=& -m^p -h_0(4 n_{++}-m-1),\label{eqn:eperspin}
\end{eqnarray}
where we have denoted $N_{++}/N=n_{++}$ and used $\sum S_i/N=m$ and the relations
\begin{eqnarray} \label{eqn:relations}
&&n_{++}+n_{+-}=\frac{1}{2}(1+m)\\
&&n_{-+}+n_{--}=\frac{1}{2}(1-m)\\
&&n_{++}+n_{-+}=\frac{1}{2}\\
&&n_{+-}+n_{--}=\frac{1}{2}.
\end{eqnarray} 
Using Stirling's formula together with \eref{eqn:eperspin} and the above relations, we can express the entropy per spin as a function of the energy and magnetization \eref{eqn:mc_s},
\begin{eqnarray}\label{eqn:mc_s_app}
s(m,\epsilon ,h_0) &=&-\left(\frac{1+m}{4}-\frac{\epsilon+m^p}{4 h_0}\right)\ln\left(\frac{1+m}{4}-\frac{\epsilon+m^p}{4 h_0}\right) \nonumber \\ 
&&- \left(\frac{1+m}{4}+\frac{\epsilon+m^p}{4h_0}\right)\ln\left(\frac{1+m}{4}+\frac{\epsilon+m^p}{4h_0}\right) \nonumber \\
&& -\left(\frac{1-m}{4}-\frac{\epsilon+m^p}{4h_0}\right)\ln\left(\frac{1-m}{4}-\frac{\epsilon+m^p}{4h_0}\right) \nonumber \\ 
&& - \left(\frac{1-m}{4}+\frac{\epsilon+m^p}{4h_0}\right)\ln\left(\frac{1-m}{4}+\frac{\epsilon+m^p}{4h_0}\right)-\ln 2.
\end{eqnarray}
The first derivative with respect to the magnetization is
\begin{eqnarray}\label{eq:dsdm}
\frac{\partial s}{\partial m} =&&-\frac{1}{4}\left(1-\frac{pm^{p-1}}{h_0}\right)\ln\frac{1}{4}\left(1+m-\frac{\epsilon + m^p}{h_0}\right) \nonumber \\ 
&&-\frac{1}{4}\left(1+\frac{pm^{p-1}}{h_0}\right)\ln\frac{1}{4}\left(1+m+\frac{\epsilon + m^p}{h_0}\right) \nonumber \\
&&-\frac{1}{4}\left(-1+\frac{pm^{p-1}}{h_0}\right)\ln\frac{1}{4}\left(1-m+\frac{\epsilon + m^p}{h_0}\right) \nonumber \\
&&-\frac{1}{4}\left(-1-\frac{pm^{p-1}}{h_0}\right)\ln\frac{1}{4}\left(1-m-\frac{\epsilon + m^p}{h_0}\right),
\end{eqnarray} 
for which $m=0$ is always a solution of $\partial s / \partial m=0$. The first derivative with respect to the energy is the generalized temperature \cite{campa},
\begin{equation}\label{eqn:beta}
\tilde \beta(\epsilon,m)=\frac{1}{4h_0}\ln\frac{(1-m-(\epsilon + m^{p})/h_0)(1+m-(\epsilon + m^{p})/h_0)}{(1+m+(\epsilon + m^{ p})/h_0)(1-m+(\epsilon + m^{p})/h_0)},
\end{equation}
from which the equilibrium temperature is obtained by setting $m=m^*$, the value of the magnetization where the entropy has its global maximum.
The second derivative of the entropy with respect to the magnetization at $m=0$ gives an indication of the order of the transition
\begin{eqnarray}\label{eq:d2sdm2}
\frac{\partial^2 s}{\partial m^2} =\quad&& p(1-p)m^{p-2}\tilde{\beta}(m,\epsilon)+\nonumber\\
&&+\frac{1}{2}\frac{(1-pm^{p-1}/h_0)(m-(\epsilon+m^p)/h_0)}{1-(m-(\epsilon+m^p)/h_0)^2} \nonumber\\
&&+\frac{1}{2}\frac{(1+pm^{p-1}/h_0)(m+(\epsilon+m^p)/h_0)}{1-(m+(\epsilon+m^p)/h_0)^2}. 
\end{eqnarray} 
Now, if we let $m\rightarrow 0$,
\begin{eqnarray}\label{eq:d2sdm2_m0}
\frac{\partial^2 s}{\partial m^2}(\epsilon,m=0) =\quad && \lim_{m\rightarrow 0} p(1-p)m^{p-2}\tilde{\beta}(m,\epsilon).
\end{eqnarray} 
For $p>2$ it is easy to see that $(\partial^2 s/\partial m^2)(m\rightarrow 0^{+})<0$ and the transition is always first order, because the paramagnetic state $m=0$ is always locally stable.

\section{Microcanonical Entropy from the Generalized Free Energy }\label{sec:mcentfromgf}
It is useful to notice that the microcanonical entropy can be obtained  by taking the supremum of the temperature derivative of the generalized free energy $f(m,\beta)$, as defined in equation \eref{eqn:cn_f}, with respect to the magnetization. This is equivalent to the procedure discussed in \cite{campa}. 
\begin{eqnarray}\label{eqn:ym1}
s_{\mathrm{mc}} (\epsilon) = \sup_{m^*}\left\{\beta^2 \frac{\partial f}{\partial \beta} ( m^* ,\beta) \right\},
\end{eqnarray}
where $\epsilon = \partial \beta f\left(m^*,\beta\right) / \partial \beta$, is the energy obtained from the generalized free energy at the value of the magnetization determined by the above condition \eref{eqn:ym1} and $m^*$ is determined by the solutions of equation \eref{eqn:sc_cn}.

The canonical entropy is calculated from the canonical free energy 
\begin{eqnarray}
&& f_{\mathrm{can}}(\beta) = \inf_{m^*} \left\{f( m^*, \beta)\right\}\\\label{eqn:ym2}
&& s_{\mathrm{can}}=\beta^2 \frac{df_{\mathrm{can}}}{d\beta} \label{eqn:ym3},
\end{eqnarray}
where $m^*$ denotes all solutions of the self-consistent equations \eref{eqn:sc_cn}. In contrast to the calculation in equation \eref{eqn:ym1}, the temperature derivative is taken $after$ the extremization with respect to the magnetization in equation \eref{eqn:ym3}.  

\begin{figure}
\centering
\includegraphics[width=\textwidth]{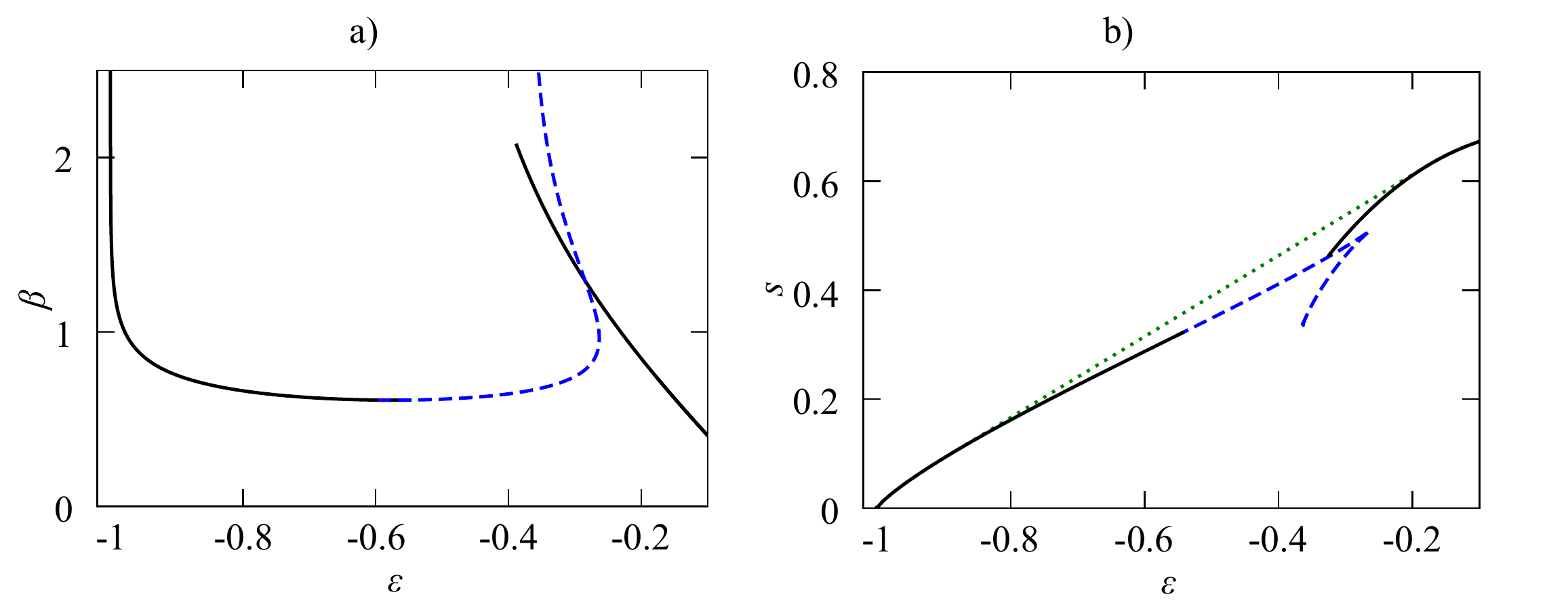} 
\caption{a) The branches of energy as obtained from the generalized free energy. Unstable solutions are in blue, the low-energy branch is the ferromagnetic and the high energy branch is the paramagnetic solution.  b) The microcanonical entropy derived from the generalized free energy is shown in black, while the green dotted line is the canonical entropy. The unstable branch is shown dashed. These branches correspond to the three possible solutions of the self-consistent equations \eref{eqn:sc_cn}.}\label{fig:ym1}
\end{figure}

In \fref{fig:ym1} a) we show the three branches of the energy as the derivative of the generalized free energy for $h_0=0.5$. The paramagnetic and ferromagnetic branches are shown in black, while the unstable solutions are shown in blue. It is not indicated where the ferromagnetic and paramagnetic branches are metastable. Comparing \fref{fig:ym1} a) to \fref{fig:p3_se} b) we note the following. When we calculate the canonical free energy, we ignore the canonically unstable (blue in \fref{fig:ym1} a)) solutions, as we look for the branch which minimizes the free energy. Such canonically unstable solutions correspond to the strictly convex part of the microcanonical entropy \cite{hull-gross}. However, part of this canonically unstable solution is stable and metastable (ferromagnetic) in the microcanonical ensemble (black and red dashed in \fref{fig:p3_se} b)). If, on the other hand, we keep track of all three branches of the generalized free energy, we can obtain the microcanonical entropy via relation \eref{eqn:ym1} as shown in \fref{fig:ym1} b). Here, the black line is the microcanonical entropy, while the green dotted line shows the canonical entropy that would have been obtained from a conventional calculation as done in equation \eref{eqn:ym3}.

\end{document}